\documentclass[%
reprint,
superscriptaddress,
showpacs,
bibnotes,
amsmath,amssymb,
pra,
floatfix,
]{revtex4-1} 

\usepackage{graphicx}
\usepackage{dcolumn}
\usepackage{bm}
\usepackage{multirow}
\usepackage{array}
\usepackage[usenames,dvipsnames]{xcolor}
\definecolor{nblue}{rgb}{0.0, 0.0, 1.0}
\usepackage[colorlinks,linkcolor=nblue,urlcolor=nblue,citecolor=nblue,plainpages=false,pdfpagelabels,breaklinks]{hyperref}
\usepackage{undertilde}
\usepackage{booktabs}
\usepackage{ctable}
\usepackage{upgreek}
\usepackage{mathrsfs}
\usepackage{amssymb}
\usepackage{amsbsy}
\usepackage{color}
\usepackage{cancel}
\usepackage[bbgreekl]{mathbbol}
\usepackage{marginnote}
\usepackage{amsfonts}
\usepackage[caption=false]{subfig}
\newcommand{\bcen}{\begin{center}}
\newcommand{\ecen}{\end{center}}
\newcommand{\btab}{\begin{tabular}}
\newcommand{\etab}{\end{tabular}}
\newcommand{\bdes}{\begin{description}}
\newcommand{\edes}{\end{description}}

\newcommand{\beq}{\begin{equation}}
\newcommand{\eeq}{\end{equation}}
\newcommand{\bea}{\begin{eqnarray}}
\newcommand{\eea}{\end{eqnarray}}
\newcommand{\non}{\nonumber}

\newcommand{\half}{\frac{1}{2}}
\newcommand{\bary}{\begin{array}}
\newcommand{\eary}{\end{array}}
\newcommand{\benum}{\begin{enumerate}}
\newcommand{\eenum}{\end{enumerate}}
\newcommand{\bitem}{\begin{itemize}}
\newcommand{\eitem}{\end{itemize}}

\newcommand{\bra}[1]{{\langle #1 |}}
\newcommand{\ket}[1]{| #1 \rangle}

\newcommand{\eqn}[1] {eqn.~(\ref{#1})}

\newcommand{\sect}[1] {Sec.~\ref{#1}}

\newcommand{\fig}[1]{fig.~\ref{#1}}

\makeatletter

\newcommand{\Rmnum}[1]{\expandafter\@slowromancap\romannumeral #1@}
\makeatother

\newlength{\myfigwidth}
\newlength{\myhalffigwidth}
\newcommand{\mysum}[1]{\displaystyle{\sum^{}_{#1}}}
\newcommand{\halfsum}[1]{\displaystyle{\sum^{\sim}_{#1}}}

\newcommand{\dW}{{\mathbb{W}}}
\newcommand{\dL}{{\mathbb{L}}}

\newcommand{\SU}[1]{SU($#1$)}

\newcommand{\mylabel}[1]{\label{#1}} 

\newsavebox{\measurebox}

\usepackage{bbm}

\usepackage{lineno}

\begin{document}




\title{Synthetic-gauge-field-induced resonances and Fulde-Ferrell-Larkin-Ovchinnikov states in a one dimensional optical lattice}
\author{Sudeep Kumar Ghosh}
\email{sudeep@physics.iisc.ernet.in}
\affiliation{Centre for Condensed Matter Theory, Department of Physics, Indian Institute of Science, Bangalore 560 012, India}
\author{Umesh K. Yadav}
\affiliation{Department of Physics, Lovely Professional University, Phagwara - 144411, Punjab, India}

\date{\today}

\begin{abstract}
Coherent coupling generated by laser light between the hyperfine states of atoms, loaded in a $1$D optical lattice, gives rise to the ``synthetic dimension'' system which is equivalent to a Hofstadter model in a finite strip of square lattice. An \SU{M} symmetric attractive interaction in conjunction with the synthetic gauge field present in this system gives rise to unusual effects. We study the two-body problem of the system using the $T$-matrix formalism. We show that the two-body ground states pick up a finite momentum and can transform into two-body resonance like features in the scattering continuum with a large change in the phase shift. As a result, even for this $1$D system, a critical amount of attraction is needed to form bound states. These phenomena have spectacular effects on the many body physics of the system analyzed using the numerical density matrix renormalization
group technique. We show that the Fulde-Ferrell-Larkin-Ovchinnikov (FFLO) states form in the system even for a ``balanced'' gas and the FFLO momentum of the pairs scales linearly with flux. Considering suitable measures, we investigate interesting properties of these states. We also discuss a possibility of realization of a generalized interesting topological model, called the Creutz ladder.
\end{abstract}


\maketitle


\section{Introduction}

Low dimensional quantum systems have been an active field of research over the last few decades marked by remarkable developments in device engineering and amazing discoveries \cite{Girvin1998,Geller2001,Mannhart2008,Neto2009,Goerbig2011}. One such example is the formation of novel states with exotic pairing \cite{Casalbuoni04,Uji2006,Gerber2014}. The Fulde-Ferrell-Larkin-Ovchinnikov (FFLO) state plays a central role in understanding such exotic pairing mechanisms and is of importance in different areas of physics \cite{Casalbuoni04}. An FFLO state \cite{Fulde64,Larkin64} is an exotic quantum phase characterized by a spatially non-uniform order parameter and finite center of mass pairing of fermions.

Ultracold atomic systems have provided an ideal platform to study the physics of strongly interacting many body systems in an unprecedentedly controlled and clean environment \cite{Bloch2012,Bloch2008}. Quantum simulation of the low dimensional systems in cold atoms has given a better understanding of the static and dynamical properties of these systems both in equilibrium and nonequilibrium \cite{Guan2013,Bloch2008}. Realization of the Tonks-Girardeau gas of hard-core bosons \cite{Paredes04} and a quantum Newton's cradle \cite{Kinoshita2006} in $1$D, Kosterlitz-Thouless transition \cite{Hadzibabic06} in $2$D are few of such examples. But in spite of extensive theoretical \cite{Parish07,Liu2007,Koponen07,Dong13,Zheng2016} and experimental \cite{Zwierlein2006,Part06,Liao2010} efforts, a direct observation of an FFLO state still remains elusive. It is hindered by technical limitations in $3$D and $2$D \cite{Zwierlein2006,Part06,Hu2006}, but the $1$D Fermi gases with population imbalance \cite{Yang2001,Orso2007,Casula08,Andreas2008,Feiguin2007,Rizzi08,Tezuka08,Feiguin2011,Guan2013} are believed to be the most suitable candidates (with already an indirect observation reported in the ref. \cite{Liao2010}). 

Gauge fields used in the gauge theories are central to the understanding of the nature of interactions between elementary particles. Cold atoms being neutral objects, gauge fields are simulated artificially and are called synthetic gauge fields \cite{Goldman2014,Dalibard2011}. There are several experimental realizations of synthetic gauge fields in cold atoms both in continuum \cite{Lin2011,Lin2009A,Lin2009B} and lattice geometries \cite{Aidelsburger2011,Miyake2013,Aidelsburger2013}. In cold atomic systems, they give rise to interesting phenomena \cite{Zhai2015,Goldman2014,Shenoy2012} such as the formation of interesting magnetic phases during the superfluid to Mott insulator transition \cite{Cole12}, generating exotic quantum phases \cite{Redic12,Sedrakyan12,Cai12}, producing fundamental changes in the two-body scattering of particles \cite{Jayantha_twobody} and discernible effects in the size and shape of a trapped cloud \cite{Sudeep2011} etc.

On the other hand, enlarged unitary symmetries such as \SU{M > 2} are crucial to the standard model of particle physics and the theory of quantum chromodynamics (QCD). For example, the physics of hedrons is described by an approximate \SU{M} symmetry group where $M$ is the number of species of quarks. But, in realistic condensed matter systems, this extended continuous symmetry is uncommon and generally introduced as a purely mathematical concept. There are, however, special cases when it emerges spontaneously, e.~g. realization of an \SU{4} Kondo effect in semiconductor quantum dots \cite{Keller2014}, \SU{4} symmetry in graphene \cite{Neto2009,Goerbig2011} and strongly correlated electrons with orbital degeneracy \cite{Kugel2015} etc. From the theoretical perspective, enlargement of the symmetry from \SU{2} to \SU{M} and doing a perturbative expansion in $1/M$ (with large $M$) have been useful in understanding the physics of Kondo lattice models \cite{Coleman1983}, Hubbard model with extended symmetry \cite{Read89,Affleck1988} etc. Ultracold atoms loaded in optical lattices \cite{Bloch2008,Bloch2012,lewenstein2012} provide natural realizations of strongly correlated many body fermionic systems with extended \SU{M > 2} symmetry. Indeed, there are several similarities between the ultracold atomic systems with \SU{M > 2} symmetry and dense QCD matter at low temperatures \cite{Rapp07,Rapp08,Maeda2009}. Remarkable recent developments \cite{Fukuhara07,Tey2010,DeSalvo10,Taie12,Cazalilla14,Stellmer2014} have made it possible to realize several such systems in cold atoms with controlled interactions and to study their interesting behaviors. Alkaline earth atoms are generic candidates for such realizations due to their special properties \cite{Cazalilla14}. Realizations of \SU{6} symmetric systems using $^{173}$Yb \cite{Fukuhara07,Taie12,Sugawa2013} and \SU{10} symmetric systems using $^{87}$Sr \cite{Tey2010,DeSalvo10,Stellmer2014} are such examples.

Hence, being naturally motivated, we consider a multicomponent $1$D system with \SU{M} symmetric attractive interaction and synthetic gauge fields in this article and show that the FFLO states can be realized in this system even without any ``population imbalance'' between the flavors. The system under consideration is a recent realization of the Hofstadter model \cite{Hofstadter} in a finite strip of square lattice with a system of atoms having multiple hyperfine states loaded in a $1$D optical lattice. The hyperfine states provide an additional dimension, called the ``synthetic dimension'' (SD). Raman assisted coherent coupling between the hyperfine states using laser light generates tunneling along this synthetic dimension. This system has received a large recent experimental \cite{Mancini15,Stuhl2015} and theoretical \cite{Celi2014,Zeng2015,Simone2015,Sudeep2015,Zhongbo2015} attention. It was shown that the non-interacting SD system itself displays rich physics like the formation of chiral edge states and produces a synthetic Hall ribbon \cite{Celi2014,Mancini15,Stuhl2015}. 

The experimental realizations of the SD system are most naturally possible in systems which also have \SU{M} symmetric interactions between the flavors. In these systems, the \SU{M} symmetric interaction manifests itself as ``long-ranged'' along the synthetic direction but is of ``contact type'' in the physical direction. Previous studies \cite{Rapp07,Capponi08,Klingschat10,Pohlmann13} of $M$ flavor fermions in $1$D with \SU{M} symmetric attractive interactions and without synthetic gauge fields revealed the formation of \SU{M} singlet bound states (``baryons'') and their quasi-long-range color superfluidity. With the synthetic gauge fields, as in the SD system, recent studies showed that these baryons get squished \cite{Sudeep2015} and form novel squished-baryon quasi-condensates \cite{Sudeep2016b}. Also, the SD system with repulsive \SU{M} symmetric interaction has been shown to be interesting both for bosonic \cite{Bilitewski2016} and fermionic particles \cite{Zeng2015,Simone2015,Zhongbo2015}. 

In this article, we explore the rich physics of the SD system with \SU{M} symmetric attractive interaction following the didactic route of performing a Bardeen-Cooper-Schrieffer (BCS) like analysis: first we consider the two-body instabilities of the system and then we look at their effects in the many body setting. We ask the question: ``What are the novel effects brought solely by the synthetic gauge field in this interacting SD system?'' We show that the synthetic gauge fields along with the \SU{M} symmetric interaction cause unusual effects both in two-body and many-body physics of this system. At the two-body level, two-body bound states (dimers) can form only in some regime of total center of mass momentum (COM) and the strongest dimers have finite COM scaling linearly with the flux ($\phi$). One important spin-off of our two-body analysis is that these dimers can transform into two-body resonance like features in the scattering continuum over a range of COM solely due to finite $\phi$ and gives a large change in the phase shift. These unusual phenomena have interesting consequences in the many-body physics of the system which we investigate using the numerical density matrix renormalization group (DMRG) \cite{White92,White93,Schollw2005,Schollwöck201196} method. Due to the formation of finite momentum dimers, FFLO states are stabilized in the system even with no ``imbalance'' between different flavors. We also point out that these FFLO correlations get suppressed with decreasing the strength of the interaction and can give rise to strongly interacting normal states due to the presence of the resonance like features in the two-body sector. Finally, we discuss a possible realization of the Creutz ladder model \cite{Creutz94} in this system.

This article is organized as follows. We delineate the model under consideration in \sect{sec:model} and discuss the single particle spectrum of the system in the \sect{sec:sing_part}. The two-body physics of the system is examined in \sect{sec:two_body} and \sect{sec:many_body} contains an analysis of the many body physics using DMRG. Finally, we give a summary of the results and an outlook in \sect{sec:summary}.

\section{Model}
\mylabel{sec:model}
 For an $M$ component SD system, the hyperfine states are labeled by $\sigma = 1, \ldots, M$ (called the ``synthetic direction'') and the sites of the $1$D optical lattice are labeled by $i =1, \ldots, L$, with $L$ being the total number of sites (called the ``physical direction''). The position of a physical site is thus $x_i = i d$, where $d$ is the lattice spacing. The Raman transitions generate position dependent phase factors in the couplings along the synthetic direction. Therefore, going around a plaquette as $(i,\sigma) \rightarrow (i+1,\sigma) \rightarrow (i+1,\sigma+1) \rightarrow (i,\sigma+1) \rightarrow (i,\sigma)$, gives rise to a flux $\phi$ per plaquette which depends on the wave vector of the two Raman lasers and can be tuned by changing the angle between them \cite{Celi2014,Mancini15,Stuhl2015}. The model Hamiltonian (${\cal H}$) of the SD system interacting via an \SU{M} symmetric attractive interaction thus consists of two parts: the kinetic energy ${\cal H}_0$ and the interaction energy ${\cal H}_I$. Then we have 
\bea
{\cal H} & = & {\cal H}_0 + {\cal H}_I \,, \mylabel{eqn:total_ham}\\
{\cal H}_0 & = & {\cal H}^1_0 + {\cal H}^2_0 \,,\mylabel{eqn:sing_part_ham} \\
{\cal H}^1_0 & = & -t \sum_{i} \sum_{\sigma = 1}^{M} \left( C^\dagger_{(i+1), \sigma} C_{i, \sigma} + \mbox{h.~c.}\right) \,, \\
{\cal H}^2_0 & = & \sum_{i} \sum_{\sigma = 1}^{M-1} \left(\Omega_{i,\sigma} C^\dagger_{i,(\sigma+1)} C_{i,\sigma} + \mbox{h.~c.} \right) \,, \\
{\cal H}_I & = & -\frac{U}{2} \sum_{i, \sigma, \sigma'} C^\dagger_{i, \sigma} C^\dagger_{i, \sigma'} C_{i, \sigma'} C_{i, \sigma} \,.
\eea  
The operator $C_{i,\sigma}$ ($C^\dagger_{i,\sigma}$) annihilates (creates) a particle at a site $(i,\sigma)$ of the synthetic lattice and obeys anti-commutation (commutation) relations for fermionic (bosonic) particles. The two contributions to the single particle kinetic energy operator ${\cal H}_0$ are: i) the nearest-neighbor (n.n) tunneling Hamiltonian ${\cal H}^1_0$ along the different sites of the optical lattice with n.n tunneling amplitude $t$ and ii) the hopping Hamiltonian ${\cal H}^2_0$ along the synthetic dimension with the tunneling coefficients $\Omega_{i,\sigma}$. These coefficients have the form $\Omega_{i,\sigma}=e^{j \phi x_i}\Omega_{\sigma}$ where $j=\sqrt{-1}$ and the parameters $\Omega_{\sigma}$ depend on the details of the system. We consider $\Omega_{\sigma} = \Omega f_\sigma$ corresponding to the experimental realizations \cite{Stuhl2015,Mancini15} of the SD system. Here, $f_\sigma = \sqrt{F(F+1)-(F-\sigma+1)(F-\sigma)}$ with $F = (M-1)/2$ being the total spin of the atoms. The position dependent phase in $\Omega_{i,\sigma}$ generates the necessary Peierls phase for producing flux $\phi$ per plaquette in the optical lattice. The \SU{M} symmetric two-body attractive interaction ${\cal H}_I$ has strength $U$ ($>0$). It is of ``contact type'' in the physical direction and is ``long-ranged'' along the synthetic direction, enabling any two hyperfine states to interact with the same strength $U$.

\begin{figure}[!t]
\centerline{\includegraphics[width=1.1\myfigwidth]{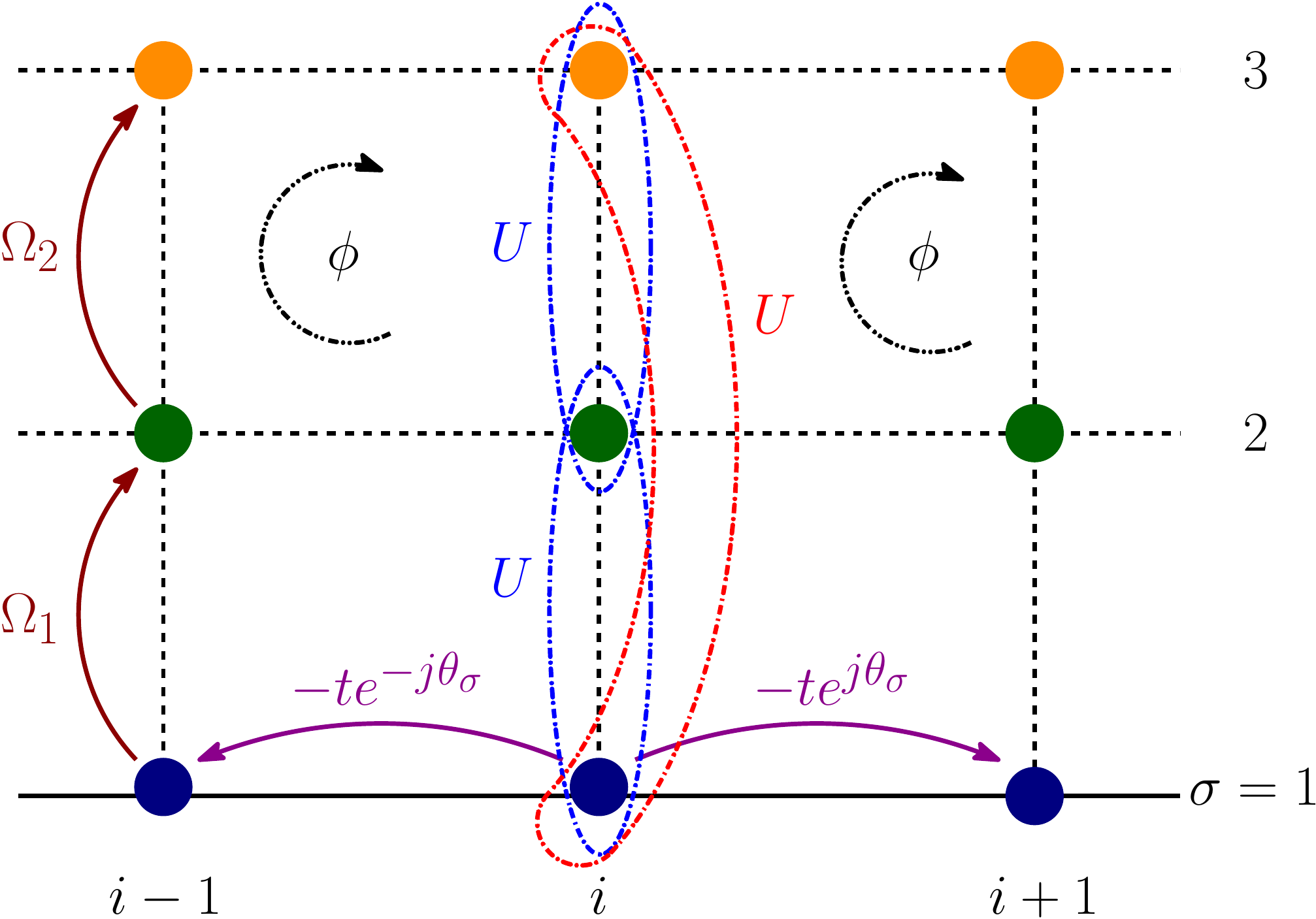}}
\caption{(Color online) Schematic plot of the synthetic dimension system with atoms in the $F = 1$ ($M = 3$) ground state manifold in the transformed basis. The ``physical sites'' of the $1$D optical lattice are labeled by $i$ and the hyperfine states are labeled by $\sigma = 1, 2$ and $3$. Tunneling along the physical direction is with an amplitude $t$ and a phase factor $\theta_\sigma$ (here, $j=\sqrt{-1}$). The spin flip hopping along the synthetic direction accompanies an amplitude $\Omega_\sigma$ in going from $\sigma \rightarrow \sigma+1$ at a particular physical site $i$. Thus moving around a plaquette of this synthetic ladder gives rise to a flux $\phi$. The \SU{M} symmetric two-body interaction is of strength $U$ and is ``long-ranged'' in the synthetic direction shown by the dashed-dotted curves.}
\mylabel{fig:schematic_plot}
\end{figure}

The physics of the SD system is more conveniently described in a different basis generated by using a local unitary transformation \cite{Sudeep2015}. This transformation, $b^{\dagger}_{i,\sigma} = e^{-j \theta_\sigma x_i} C^{\dagger}_{i,\sigma}$, creates the new operators $b^\dagger_{i,\sigma}$ which obey  same anti-commutation or commutation relations as the $C^\dagger_{i,\sigma}$ operators. In this transformed basis, different terms of the Hamiltonian ${\cal H}$ (\eqn{eqn:total_ham}) become
\bea
{\cal H}^1_0 & = & -t \sum_{i} \sum_{\sigma = 1}^{M} \left(e^{j \theta_\sigma} b^\dagger_{(i+1), \sigma} b_{i, \sigma} + \mbox{h.~c.}\right) \,\,,\mylabel{eqn:Ht} \\
{\cal H}^2_0 & = & \sum_{i} \sum_{\sigma = 1}^{M-1} \Omega_\sigma \left(b^\dagger_{i,(\sigma+1)} b_{i,\sigma} + \mbox{h.~c.} \right) \,\,, \mylabel{eqn:HOmg} \\
{\cal H}_I & = & -\frac{U}{2} \sum_{i, \sigma, \sigma'} b^\dagger_{i, \sigma} b^\dagger_{i, \sigma'} b_{i, \sigma'} b_{i, \sigma} \,\,. \mylabel{eqn:Hint}
\eea  
Here, the phase factor $\theta_\sigma = (\sigma -1)\phi$. Interestingly, we note that in this transformed basis the position dependence of the tunneling along the synthetic dimension is suppressed (\eqn{eqn:HOmg}) at the cost of putting a position independent phase factor in the tunneling along the physical direction (\eqn{eqn:Ht}). The SD system in this basis for $M=3$ is schematically depicted in the \fig{fig:schematic_plot}. Throughout this article, we consider this basis and work with the units of $\hbar$ and $d$ being unity.

\begin{figure}[!b]
\centerline{\includegraphics[width=1.1\myfigwidth]{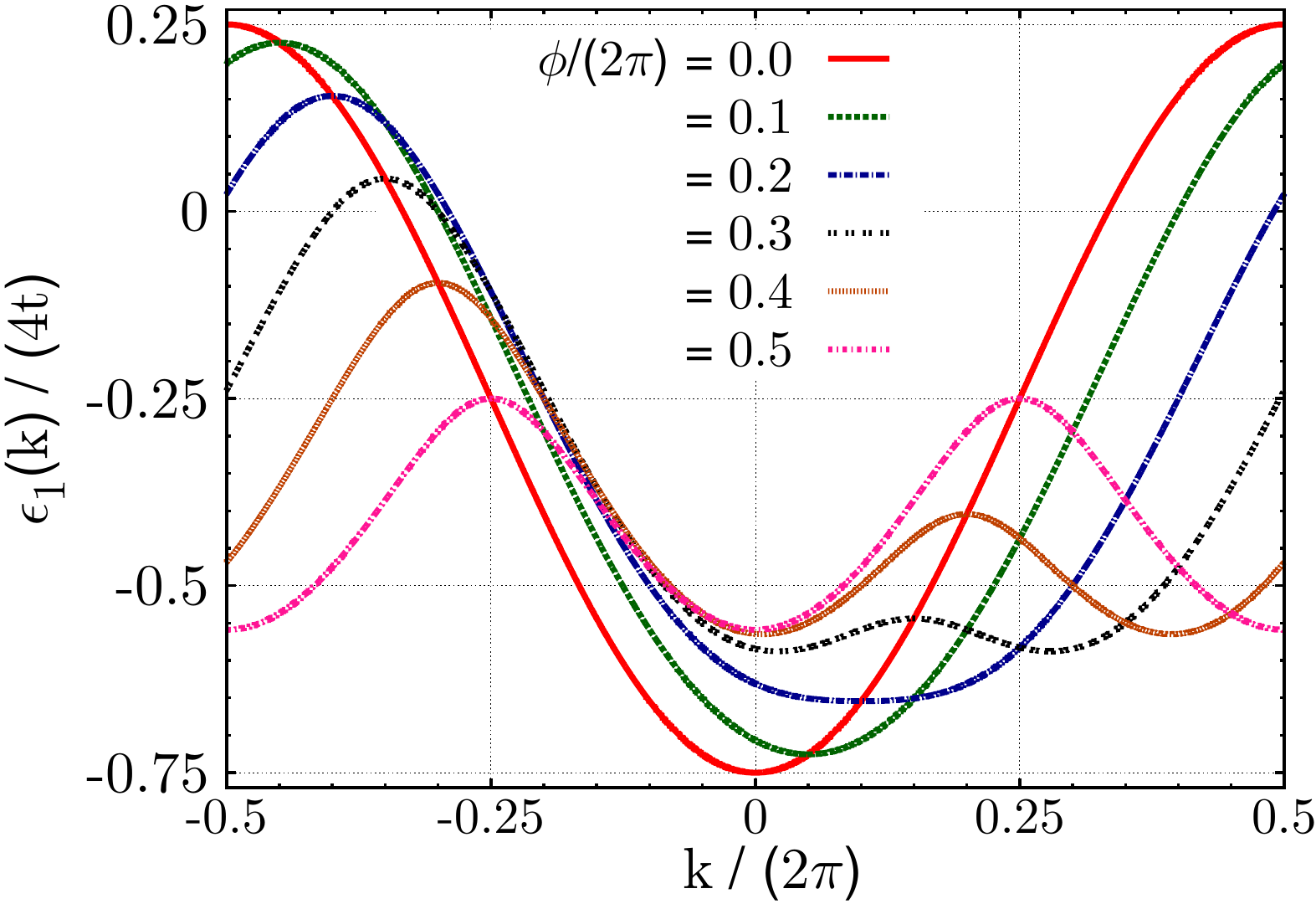}}
\caption{(Color online) Single particle dispersion of the first band for the $M=2$ SD system with $\Omega/t = 1$ and different values of $\phi$.}
\mylabel{fig:bands}
\end{figure}

\section{Single particle physics}
\mylabel{sec:sing_part}
We consider periodic boundary condition (PBC) in the physical direction and take momentum as a good quantum number. The single particle kinetic energy operator ${\cal H}_0$ in momentum space can be rewritten in the form
\bea
{\cal H}_0 & = & \sum_{k}\sum_{\sigma = 1}^{M} \varepsilon_\sigma(k) b^{\dagger}_{k,\sigma} b_{k,\sigma} \non\\
&+& \sum_{k}\sum_{\sigma = 1}^{M-1} \Omega_\sigma \left(b^\dagger_{k,(\sigma+1)} b_{k,\sigma} + \mbox{h.~c.} \right) \;. \mylabel{eqn:H0_kspace}
\eea
Here, we have defined $\varepsilon_\sigma(k) = -2 t \cos(k-\theta_\sigma)$ and $b^{\dagger}_{k,\sigma} = 1/\sqrt{L} \sum_i e^{j k x_i} b^{\dagger}_{i,\sigma}$ with $L$ being the total number of physical sites. We note that the first term in \eqn{eqn:H0_kspace} describes the spin-orbit coupling generated by the synthetic gauge field and the second term acts as the Zeeman term with Zeeman field strength $\Omega$. In the limit of $\Omega \to 0$ \footnote{The limit $\Omega = 0$ and $\phi \neq 0$ is not well defined because strictly speaking the unitary transformation to change the basis is valid only when $\Omega \neq 0$.}, the single particle dispersions have minima at $k=\theta_\sigma$ and these bands are split from each other with increasing $\Omega$. Now, using a unitary transformation ${\cal H}_0$ can be diagonalized as 
\beq
{\cal H}_0 = \sum_{k,\alpha} \epsilon_\alpha(k) a^{\dagger}_{k,\alpha} a_{k,\alpha} \;,
\eeq
where, $\epsilon_\alpha(k)$ are the energies of the single particle states labeled by $\alpha$. The unitary transformation is given by $b^{\dagger}_{k,\sigma} = \sum_{k_1,\alpha_1} R_{\sigma,\alpha_1}(k,k_1)a^{\dagger}_{k_1,\alpha_1}$ with $R(k,k_1)$ being a unitary matrix which is diagonal in the momentum indices, i.~e. has the form $R_{\sigma,\alpha}(k,k_1) = R_{\sigma,\alpha}(k) \delta_{k,k_1}$. For the particular case of $M=2$ analytical solutions of the single particle band structure are possible and they are given by
\beq
\epsilon_\alpha(k) = \frac{\varepsilon_1(k)+\varepsilon_2(k)}{2} + (-1)^\alpha \sqrt{\Omega^2 + \left(\frac{\varepsilon_1(k)-\varepsilon_2(k)}{2}\right)^2} , \non
\eeq
with $\alpha = 1$ and $2$. We note that for this case at a particular $\Omega$, there is an interesting change in the single particle spectrum of the system with changing $\phi$ and the lowest band gradually develops a double well structure as shown in \fig{fig:bands}. 

\section{Two-body physics}
\mylabel{sec:two_body}
In this section, we investigate the physics of two particles interacting via ${\cal H}_I$ (\eqn{eqn:Hint}) in the SD system.  To proceed, we recast ${\cal H}_I$ in the momentum space as
\beq
\mylabel{eqn:int_mom}
{\cal H}_I = -\frac{U}{2 L} \sum_Q \sum_{\sigma \sigma'} P^\dagger_Q(\sigma, \sigma') P_Q(\sigma, \sigma') \,,
\eeq
where, $Q$ is the total canonical COM of a pair created by the pair creation operator 
\beq \mylabel{eqn:pair_create}
P^\dagger_Q(\sigma, \sigma') = \sum_{k} b^\dagger_{\left(\frac{Q}{2} + k \right), \sigma} b^\dagger_{\left(\frac{Q}{2} - k \right), \sigma'} 
\eeq
with relative momentum $k$. If $k_1$ and $k_2$ are the individual momenta of the two particles constituting the pair then $Q \equiv (k_1+k_2)$ and $k \equiv \frac{k_1-k_2}{2}$.  We now use the $T$-matrix formulation to analyze the two-body problem.

\begin{figure*}[!t]
\centerline{\includegraphics[width=2.32\myfigwidth]{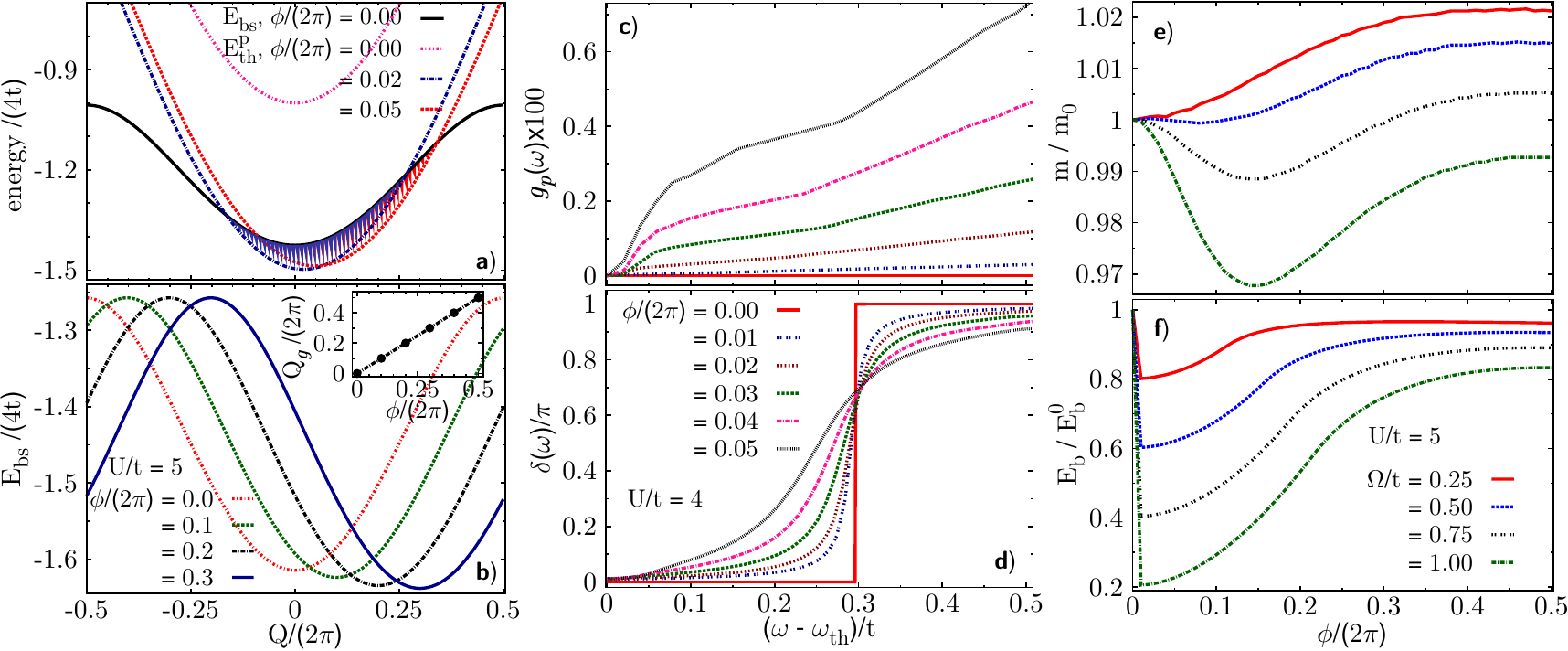}}
\caption{(Color online) Two-body bound state properties of a fermionic SD system with $M = 2$. Panels (a) and (b) respectively show the variations of the pairing threshold ($E^p_{th}$) and the bound state energy ($E_{bs}$) as a function of the canonical COM $Q$ with $\Omega/t = 1$. The minimum of $E^p_{th}$ (see (a)) and $E_{bs}$ (see (b)) occur at $Q=Q_g$ and it scales as $Q_g \propto \phi$ (inset of (b)). We note from (a) that $E_{bs}(Q)$ of $\phi = 0$ (which is with $U/t=4$) can be just above $E^p_{th}(Q)$ of $\phi \neq 0$ in some regime of $Q$ (hatched regimes below the black curve). In this situation, the fate of the bound states is explored in (c) and (d) focusing at $Q = 0$. Panels (c) and (d) respectively show the total PDOS ($g_p(\omega)$) and the phase shift ($\delta(\omega)$) as a function of energy ($\omega$) with $\Omega/t = 1$. Finally, in panels (e) and (f), we show as a function of $\phi$ the behaviors of the mass ($m$) and the binding energy ($E_b$) respectively for the strongest bound state occurring at $Q=Q_g$ with $U/t = 5$ (which is larger than $U/t = 4$ at which resonances occur). Here, $m_{0}$ and $E^{0}_{b}$ are the values of $m$ and $E_b$ respectively for a $1$D free Fermi gas ($\phi = \Omega=0$).}
\mylabel{fig:bs_prop}
\end{figure*}

\subsection{Formulation of the two-body problem} 
We define a two-body state as $\ket{K} \equiv \ket{k_1,\alpha_1;k_2,\alpha_2} = a^\dagger_{k_1,\alpha_1} a^\dagger_{k_2, \alpha_2} \ket{0}$, where $\ket{0}$ is the vacuum state. It is noted that the kinetic energies of the states $\ket{K}$ and $\ket{\tilde{K}} \equiv \ket{k_2, \alpha_2; k_1, \alpha_1} = \zeta \ket{K}$ (the parameter $\zeta = -1$ ($+1$) for fermions (bosons)) are the same. The linearly independent states are, therefore, with $k_1 \ge k_2$ and we define a sum only over these states as $\halfsum{K} \equiv \sum_{k_1 \ge k_2, \alpha_1,\alpha_2}$. The non-interacting two particle spectrum $E(K)$ corresponding to the state $\ket{K}$ is $E(K) = \left[\epsilon_{\alpha_1}(k_1) + \epsilon_{\alpha_2}(k_2)\right]$. Using the $T$-matrix formulation, described in detail in the Appendix, we now investigate the bound state properties of the system. The effective scattering potential (\eqn{eqn:eff_sca_pot}) coming from the interaction term (\eqn{eqn:interaction_eff}) acts over all the $M^2$ scattering channels of the two-body system but symmetry properties of the two-body wave function forces only $\kappa = (M(M+\zeta))/2$ of them to be truly independent. We determine these $\kappa$ number of bound states with energies $E_{bs}(Q)$ by solving for the poles of the $T$-matrix (i.~e. \eqn{eqn:inveqn}).

We define $W_c(K)$ (see \eqn{eqn:pairpot}) to be the pair amplitude corresponding to the state $\ket{K}$ at a particular channel $c$.
Then, we can define the pair density of states (PDOS) $g^{p}_{c,c_1}(\omega)$, which measures the propensity of bound state formation in the system, corresponding to an incoming state at channel $c$ and an outgoing state at channel $c_1$ with energy $\omega$ as
\beq
g^{p}_{c,c_1}(\omega) = \frac{\pi}{2L} \halfsum{K} W_c^{*}(K) W_{c_1}(K) \delta(\omega^{+} - E(K))\,\,.
\eeq
Bound states can now form in the system in the regime below an energy value where the PDOS is zero. This energy value defines the pairing threshold $E^{p}_{th}(Q)$ of the  system, i.~e. $E_{th}^p (Q) = min_{\{c,c_1\}} \omega_{c,c_1}$, where $\omega_{c,c_1}$ is the lowest value of $\omega$ in a particular $(c,c_1)$ for which the PDOS, $g^{p}_{c,c_1}(\omega) = 0^{+}$. Hence, the pairing threshold measures the threshold energy for bound state formation, i.~e. a two-body state with energy value less than $E^p_{th}$ can form a bound state pair while that with energy greater than $E^p_{th}$ goes into the scattering continuum. The binding energy $E_{b}(Q)$ of a bound state with energy $E_{bs}(Q)$ can now be defined with respect to the $E^{p}_{th}(Q)$ as $E_b(Q) = [E^{p}_{th}(Q) - E_{bs}(Q)]$. We can also define another threshold, known as the two-body threshold, which is the minimum energy of the non-interacting two particle spectrum, i.~e. $E_{th}(Q) = min_{K} E(K)$. Interestingly, in general $E_{th} (Q) \leq E^p_{th}(Q)$. In the following, we are also interested to look into the behavior of the mass $m(Q)$ of a bound state which is defined as
\beq
 m^{-1}(Q) = \frac{\partial^2 E_{bs}(Q_1)}{\partial {Q_1}^2}\bigg|_{Q_1=Q}.
\eeq
 
\subsection{Results of the two-body problem}
\mylabel{sec:two_body_results}
The results of the two-body problem, obtained using the formalism just discussed, are presented here. In the limit of $\phi = 0$, exact analytical form of the secular matrix (see the Appendix) can be obtained and exact forms of different bound state properties can be found. If the bound states are labeled by an integer function $s(\alpha_1,\alpha_2)$ which takes values $1, 2, ..., \kappa$, then they have energy $E_{bs}^{s} = \left[ - \sqrt{U^2 + 16 t^2 \cos^2(Q/2)} + \Omega X_{s} \right]$ with $X_{s(\alpha_1,\alpha_2)} = [(2\alpha_1-M-1)+(2\alpha_2-M-1)]$. The allowed values of $(\alpha_1,\alpha_2)$ are determined by the statistics obeyed by the particles. The pairing threshold is then $\varepsilon^{p0}_{th} = - 4t \cos(Q/2) + \Omega X_{0}$ with $X_0$ being the value of $X_{s(\alpha_1,\alpha_2)}$ corresponding to $(\alpha_1,\alpha_2) = (1,2)$ for fermions and $(1,1)$ for bosons. The mass of the bound states (independent of $\Omega$) has a simplified form for $Q=0$ given by $m(0) = \sqrt{U^2 + 16 t^2}/(4 t^2)$. And, for $U \gg t$, $m(0) \sim U/(4 t^2)$ which can be understood by noting that in this limit particles hop to their neighboring sites via virtual processes with a kinetic energy gain $\sim 4t^2/U$.

We now consider the effect of finite flux on the bound states and concentrate only on the fermionic case. A similar analysis can be readily adopted for bosonic particles. The single particle SD system with finite flux itself is very rich \cite{Celi2014,Mancini15,Stuhl2015} and an additional \SU{M} symmetric interaction brings in non-trivial effects noticed in the refs. \cite{Sudeep2015,Sudeep2016b,Zeng2015,Simone2015,Bilitewski2016}. Hence, we expect qualitative changes in the two-body bound state spectrum of the system as a consequence of $\phi \neq 0$. We consider the $M = 2$ as an example and show the results (obtained numerically) in \fig{fig:bs_prop}. Similar physics is at play for other $M$ ($> 2$) systems but they have $\kappa$ ($> 1$) number of bound states.

Flux produces mixing of different $\alpha$-flavors. As a result, a two-body state can now be comprised of same two $\alpha$-states (which is not the case for $\phi = 0$ due to Pauli blocking) since it has a non-zero pair amplitude. This results in a sudden change in the $E^p_{th}$ from  $E^p_{th} > E_{th}$ for the $\phi=0$ case to $E^p_{th} = E_{th}$ for the $\phi \neq 0$ case. It is evident from \fig{fig:bs_prop}(a) that even for very small $\phi$ this discontinuity takes place. Another pertinent feature brought by the synthetic gauge field, which is seen both in \fig{fig:bs_prop}(a) and \fig{fig:bs_prop}(b), is that in the presence of finite $\phi$, the minima of $E^p_{th}(Q)$ and $E_{bs}(Q)$ shift to a finite value of $Q = Q_g$. This implies that the strongest bound states of the system are finite momentum dimers and they form in spite of an ostensible momentum conserving interaction term (\eqn{eqn:int_mom}). Also, as shown in the inset of \fig{fig:bs_prop}(b), $Q_g$ scales linearly with $\phi$. This linear scaling can be understood from the behavior of the lowest single particle band by looking at \fig{fig:bands}. We noted that its single well structure centered around momentum $k=\phi$, with increasing $\phi$, gradually changes to a double well structure with the two wells centered around $k=0$ and $\phi$. Then, the attractive interaction generates the strongest bound state with pairs formed from two single particle states having the lowest energy. This leads to the formation of the strongest dimers having a finite COM which scales linearly with $\phi$. The $E_{bs}(Q)$ is then symmetric around the momentum $Q_g$ of the dimers. Previous studies in $3$D spin-orbit coupled Fermi gases with detuning and Zeeman field found similar results attributed to the broken Galilean invariance of the system \cite{Lin13}. These finite momentum bound states have interesting consequences in the many body setting discussed in the next section.

The discontinuity in the $E^p_{th}$ as a function of $\phi$ can give rise to a situation when the $E_{bs}(Q)$ for $\phi = 0$ (denoted by $E^{0}_{bs}(Q)$) is above the $E^p_{th}(Q)$ for $\phi \neq 0$. In this case, an interesting phenomenon can take place in a regime of $Q$ where $E^p_{th}(Q) < E^{0}_{bs}(Q)$ (shown by the hatched regimes below the black curve in \fig{fig:bs_prop}(a)). We look into this situation a bit more closely by considering the $Q=0$ case in \fig{fig:bs_prop}(c) and \fig{fig:bs_prop}(d). In \fig{fig:bs_prop}(c), we show the behavior of the total PDOS $g_p(\omega)$ defined as $g_p(\omega) = \sum_{c,c_1} g^p_{c,c_1}(\omega)$. We note that a non-zero PDOS, which increases with increasing $\phi$, appears near the two-body threshold $\omega_{th}$. Also, the behavior of the PDOS where it just becomes non-zero is very different for the $\phi \neq 0$ case than that of the zero flux case for which it behaves as $\sim 1/\sqrt{16 t^2-\omega^2}$. In this regime, if a bound state exists for the $\phi = 0$ case due to the absence of any PDOS, we expect this bound state to acquire a finite lifetime as soon as $\phi$ becomes non-zero since the PDOS also becomes non-zero.

We investigate this phenomenon by calculating the phase shift $\delta(\omega)$ defined using the $T$-matrix as \cite{Taylor2006,Shenoy2013}, $\delta(\omega) = Arg[T(\omega^+)]$. From its behavior, the nature of a bound state can be deciphered. When there is a ``true'' bound state (infinite lifetime) in the system, the phase shift gives a sharp theta function change while for a resonance like feature corresponding to a bound state with finite lifetime, there is smooth but large change in the phase-shift \cite{Taylor2006,Shenoy2013}. The sharpness in the change of the phase shift is thus related to the lifetime of the bound state. In \fig{fig:bs_prop}(d), we show the behavior of $\delta(\omega)$ for different values of finite but small $\phi$. We note that there is a sharp theta function change in $\delta(\omega)$ for $\phi = 0$ but as soon as $\phi$ becomes $\neq 0$ there is a smooth but large change. Hence, the bound state of the $\phi = 0$ case no longer remains a ``true'' bound state when $\phi \neq 0$. Instead, its vestige as a bound state is manifested as a resonance like feature in the scattering continuum accompanying a smooth but large change in the $\delta(\omega)$. As $\phi$ increases, the sharpness of the resonances decreases and the finite lifetime acquired by the bound state decreases which is because the PDOS also increases correspondingly. We also note that the resonances appear at energies dependent on $Q$. Similar results are also found in $3$D Fermi gases with spin orbit coupling (SOC) \cite{Shenoy2013} and systems with narrow Feshbach resonances \cite{Chin2010}. Hence, to produce a true bound state even for this $1$D system a critical amount of attraction strength ($U_c$) is required and $U_c$ can go to zero at a finite center of mass. 

Finally, we present an analysis of the effect of the synthetic gauge field on two properties of the strongest bound state occurring at $Q = Q_g$, namely the mass ($m$) and the binding energy ($E_b$). We show the behaviors of $m$ and $E_b$ as a function of $\phi$ in \fig{fig:bs_prop}(e) and \fig{fig:bs_prop}(f) respectively for the $M = 2$ case with a larger value of $U$ than the one at which resonances occur. We note that although $m$ changes by a small amount, there is a large change in the $E_{b}$ as $\phi$ increases. Both of them decreases with the increase in $\Omega$ for fixed $\phi$. The sudden reduction in $E_b$ (see \fig{fig:bs_prop}(f)) as soon as $\phi \neq 0$ is due to the discontinuity in $E^p_{th}$ as discussed earlier (see \fig{fig:bs_prop}(a)). Keeping $\phi$ and $U$ fixed, as $\Omega$ increases, the effective hopping parameter of the system increases and this acts against bound state formation (gives reduction of the binding energy seen in \fig{fig:bs_prop}(f)). But, flux promotes bound state formation enhancing $E_{b}$ with increasing $\phi$ at a fixed $\Omega$. Hence, there is a competition between $\Omega$ and $\phi$ in forming bound states. Although we see from \fig{fig:bs_prop}(d) that the mass varies non-monotonically for ``larger'' values of $\Omega$, first it decreases and then increases with the increase in flux. Another interesting phenomenon is that for a fixed $\phi$ and $U$, when $\Omega$ is increased or for a fixed $\phi$ and $\Omega$, when $U$ is decreased then the zeros of the secular matrix can move above the scattering threshold and appear below the next scattering continuum giving rise to bound states in-between the bands.

\section{Many body physics}
\mylabel{sec:many_body}
We use the finite system DMRG \cite{White92,White93,Schollw2005,Schollwöck201196} algorithm, retaining upto $500$ truncated states per DMRG block with the maximum truncation error of $10^{-7}$, to simulate a fermionic SD system with $N$ number of particles and open boundary condition (OBC) along the physical direction. This system having $L$ physical sites and $M$ hyperfine states in the synthetic direction can then be viewed as a ``synthetic'' ladder with $M$ legs and $L$ rungs. The spin-flip term (\eqn{eqn:HOmg}) present in the Hamiltonian of the system reduces the symmetries of the problem only to the total occupation at a physical site $i$ to be conserved. The total density of particles ($n$) of the system is defined as $n=N/L$ and we consider $n \ll 1$.

For this many body SD system with the \SU{M} symmetric attractive interaction, we are now interested in looking into the non-trivial effects brought solely by the synthetic gauge field and the consequences of the novel phenomena occurring at the two-body level discussed in the previous section. To this end, we discuss our results considering the $M =2$ fermionic SD system as an example. We focus in the parameter regime where there is no ``population imbalance'' between the two legs. Here, the ``population imbalance'' should be defined carefully since the total number of particles in each of the legs is no longer conserved. We define average number of particles in the $\sigma$-th leg as $\langle N_\sigma \rangle$ = $\sum_{i} \langle n_{i,\sigma} \rangle$, with $n_{i,\sigma}$ being the number operator corresponding to the site ($i,\sigma$) of the ladder. Then the ``population imbalance'' in the system is defined by $(\langle N_1 \rangle$ - $\langle N_2 \rangle)$ and when there is no ``population imbalance'' $\langle N_1 \rangle$ = $\langle N_2 \rangle$.

\begin{figure}[!t]
\centerline{\includegraphics[width=1.1\myfigwidth]{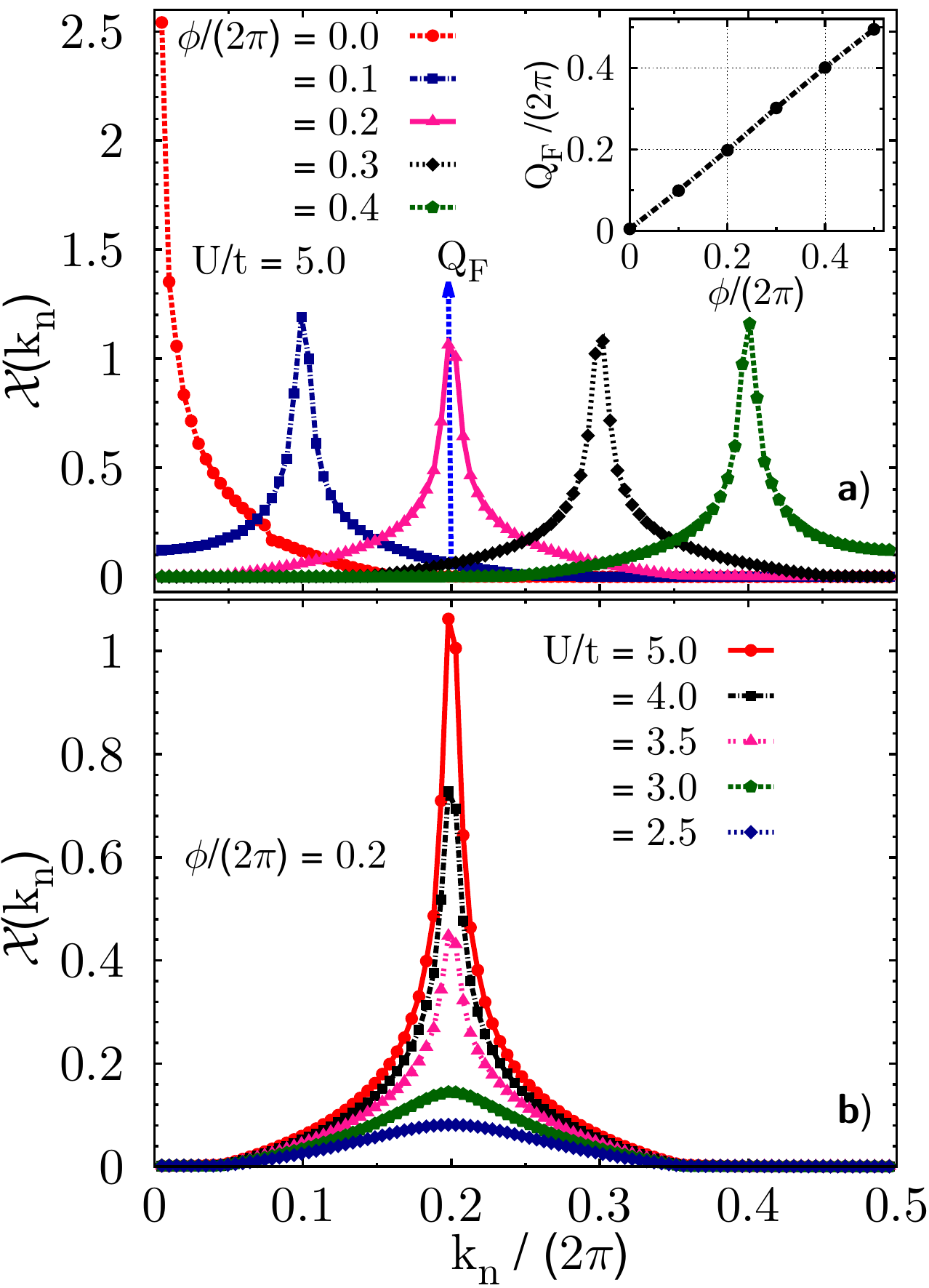}}
\caption{(Color online) \textbf{Synthetic gauge field induced FFLO states :} Variation of the pair momentum distribution function ${\cal X}(k_n)$ for the $M = 2$ (fermionic) case with $\Omega/t = 1$, $L = 100$ (number of physical sites) and total density of particles $n = 0.3$, as a function of the pair momentum ($k_n$) for different values  of $\phi$. Panel (a) shows that the peak of ${\cal X}(k_n)$ shifts to finite value of $k_n = Q_F$ as $\phi$ becomes non-zero signaling the emergence of the FFLO states. The FFLO momentum $Q_F \propto \phi$ (shown in the inset of (a)). This has its origin in the formation of finite momentum dimers shown in \fig{fig:bs_prop}(b). Panel (b) shows that at a fixed $\phi$, the FFLO peak disappears continuously as $U$ decreases.}
\mylabel{fig:FFLO}
\end{figure}

We investigate the nature of the many body ground state by computing the ground state expectation values of different local and nonlocal correlation functions of the system. Then, quasi-long-range coherence in the system can be deciphered by an algebraic decay in the non-local correlation functions. First, we consider the pair correlation function (PCF) of the system defined as
\beq\mylabel{eqn:pcf}
{\cal X}_{i,j} = \langle b^{\dagger}_{i,1} b^{\dagger}_{i,2} b_{j,2} b_{j,1} \rangle \;.
\eeq 
It measures the propensity of pair formation in the system and its algebraic decay with distance $|i-j|$ indicates the formation of a quasi-long-range pair superfluid phase such as the FFLO phase if the pairs have finite COM. We also define the pair momentum distribution function (PMF) by the Fourier transform of ${\cal X}_{i,j}$ as
\beq
{\cal X}(k_n) = \sum_{l,m} \Theta_{l}(k_n) \Theta_{m}(k_n) {\cal X}_{l,m} \,\,,
\eeq
where, $\Theta_{l}(k_n) = (2/(L+1))^\half \sin(k_n l)$ are the wave functions of a spin-less non-interacting $1$D tight binding chain with OBC, where $k_n$ takes on values $k_n = \pi n/(L+1)$ with $n = 1$, $\ldots$, $L$ and its minimum value is $k_1$. We note that the above definition of the PMF is analogous to that of the PBC in the physical direction for which it would be ${\cal X}(k) = (1/L) \sum_{l,m} e^{jk(l-m)} {\cal X}_{l,m}$. It is related to the pair creation operator (defined in \eqn{eqn:pair_create}) as ${\cal X}(k) = \langle P^{\dagger}_Q(1,2) P_Q(1,2) \rangle \delta_{k,Q}$. Hence, the PMF can be thought of being a measure of the population of pairs in the system with COM $Q$.

The results of the variations of the PMF for different values of $\phi$ are shown in \fig{fig:FFLO}(a). A narrow peak of the PMF at a finite value of $k_n > k_1$ suggests the formation of an FFLO ground state with the pairs having an FFLO momentum $Q_F$. This needs to be confirmed by comparing the algebraic decay of the FFLO correlation (\eqn{eqn:pcf}) with other correlations of the system and making sure that the FFLO correlations are indeed the dominant correlations of the system. We note from \fig{fig:FFLO}(a) that for the chosen value of $\Omega$, the ground state for the $\phi = 0$ case, is not an FFLO state ($Q_F = k_1$) and as $\phi$ deviates from zero, the $Q_F$ starts deviating from $k_1$. In addition, $Q_F$ scales linearly with $\phi$ as shown in the inset of \fig{fig:FFLO}(a). This scaling is reminiscent of and related to the scaling of the momentum ($Q_g$) of the two-body bound states shown in the inset of \fig{fig:bs_prop}(b). So, we see that the two-body finite momentum dimers (discussed in the previous section) result in the FFLO ground states in the many body SD system.

\begin{figure}[!t]
\centerline{\includegraphics[width=1.1\myfigwidth]{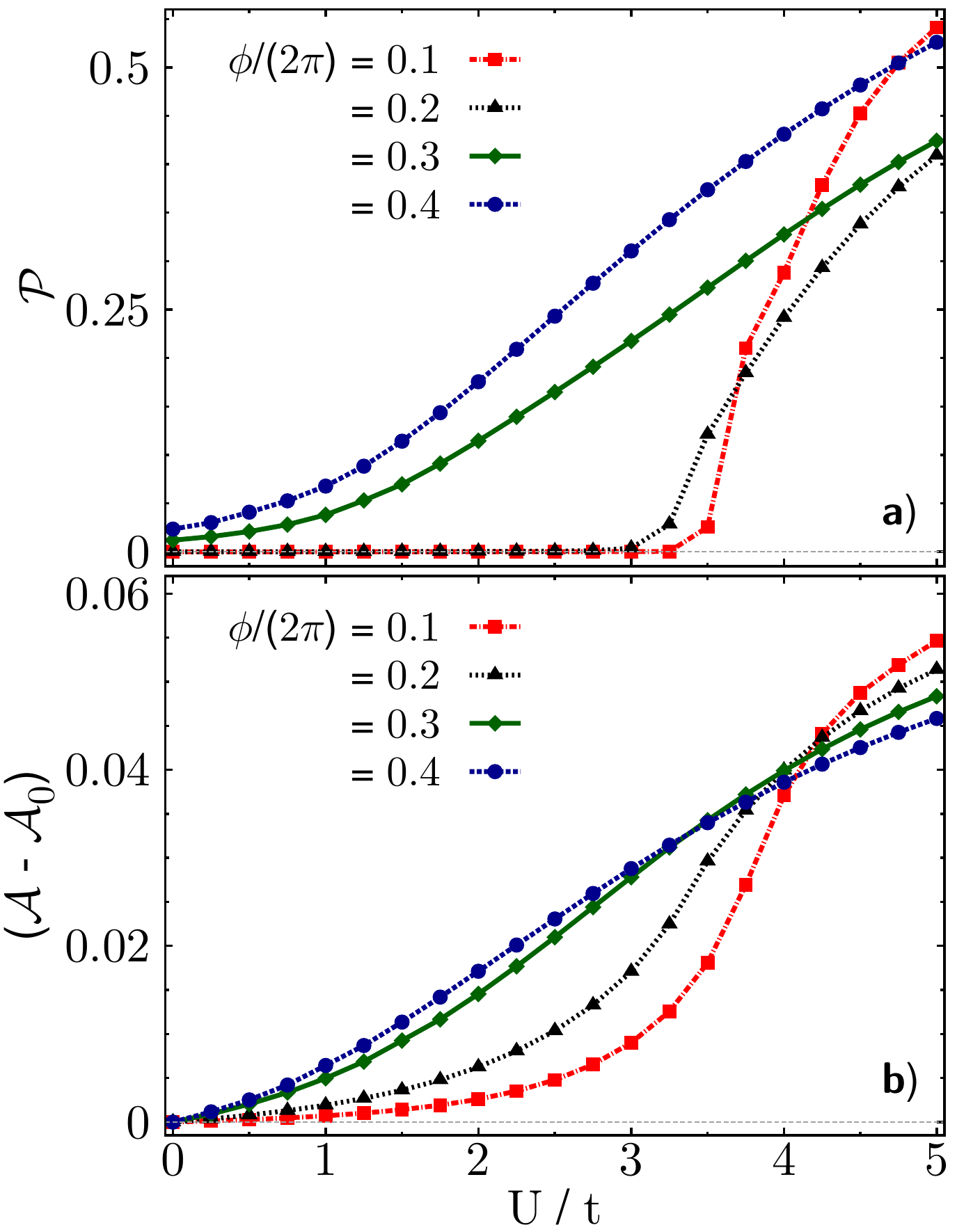}}
\caption{(Color online) Properties of the FFLO states for $M=2$, $\Omega/t=1$, $n=0.3$ and $L=100$. Panel (a) and (b) respectively show the variations of the peak anomaly (${\cal P}$) and the area (${\cal A}$) under the ${\cal X}(k_n)$ curve as a function of $U$ for different values of $\phi$. Here, ${\cal A}_0$ is the value of ${\cal A}$ at $U=0$. }
\mylabel{fig:FFLO_prop}
\end{figure}

As discussed in \sect{sec:sing_part} and shown in \fig{fig:bands}, there is a change in the single particle spectrum of the system with changing $\phi$ at a fixed value of $\Omega$. As a result, corresponding to a fixed density of particles in the system, there is a change in the topology of the Fermi surface, so called Lifshitz transition \cite{Lifshitz30}, as a function of $\phi$. The Fermi surface changes from having $2$ Fermi points to $4$ Fermi points with increasing $\phi$. We then expect to see changes in the formation of the FFLO states in the system due to this Lifshitz transition. When there are $4$ Fermi points in the system (as is the case for $\phi = 0.3$ and $0.4$ shown in \fig{fig:FFLO} with the given density), the non-interacting system ($U=0$) itself shows a sharp peak in the PMF at $Q_F \propto \phi$ although the peak value is very small compared to the one shown in  \fig{fig:FFLO}(a). But, the FFLO correlations in the system are short-ranged and there is no quasi-long range order in the system. Hence, for these cases a careful diagnosis for the FFLO states is necessary and must be done as usual by first noting a sharp peak in the PMF as well as making sure that FFLO correlations are dominant correlations of the system.

We further analyze the properties of the FFLO ground states by investigating the behavior of the ${\cal X}(k_n)$ at a fixed $\phi$ as a function of $U$ (shown in \fig{fig:FFLO}(b)). It is noted that the strength of the FFLO peak gets suppressed strongly with decreasing $U$. Finally, with continuous decrease in $U$, the peak diminishes and gets transformed into a broad hump (for the case shown in \fig{fig:FFLO}(b)) or a strongly suppressed peak (for the cases of $\phi = 0.3$ and $0.4$ having $4$ Fermi points in their respective Fermi surfaces) corresponding to a ground state with no quasi-long-range order. Hence, there is quasi-long-range coherence in the system only for $U > U_c$, where $U_c$ is a critical value of attraction. This is similar to the usual $1$D Fermi gas with Zeeman field and no spin orbit coupling \cite{Feiguin2011}. We also note that this phenomenon is consistent with our discussion of the two-body problem (the two-body bound states can form only above a critical value of attraction). 

\begin{figure}[!b]
\centerline{\includegraphics[width=1.08\myfigwidth]{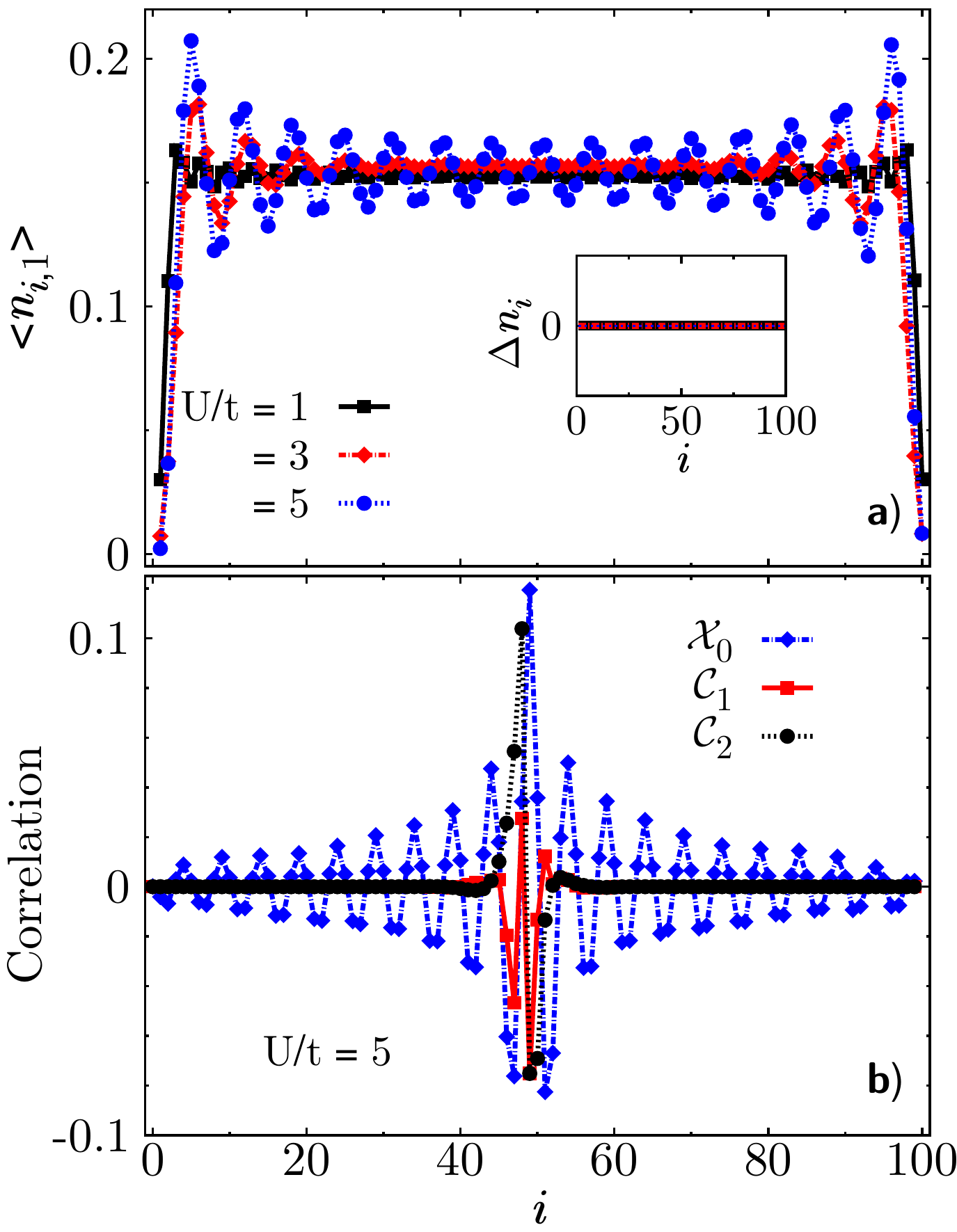}}
\caption{(Color online) \textbf{Local and non-local correlation functions :} Parameters are $M=2$, $\Omega/t=1$, $\phi/(2\pi) = 0.2$, $n=0.3$ and $L=100$. a) Variations of the local average density of particles $\langle n_{i,1} \rangle$ in the lowest $\sigma = 1$ leg as a function of the site number $i$ are shown for two different values of $U$. In the inset, we show the variation of the difference $\Delta n_i = \langle n_{i,1} \rangle - \langle n_{i,2} \rangle$ for the same $U$ values. b) Behaviors of the non-local correlation functions ${\cal X}_0$, ${\cal C}_1$ and ${\cal C}_2$ (defined in the text) with respect to the central site at $L/2$ are shown. We note that the FFLO correlation (${\cal X}_0$) is the slowest to decay.}
\mylabel{fig:Corr_prop}
\end{figure}

To get a better understanding of this phenomenon of vanishing of FFLO correlations with decreasing $U$, we define the following two properties of the FFLO peaks shown in \fig{fig:FFLO_prop}. 1) The peak anomaly (${\cal P}$) \cite{Rizzi08} defined as, ${\cal P} = \left[ 2{\cal X}(Q_F) - {\cal X}(Q_F + k_1) - {\cal X}(Q_F - k_1) \right]$. It can be thought to be proportional to the difference in the right and left discrete derivatives of ${\cal X}(k_n)$ evaluated at $k_n = Q_F$. It measures the anomaly of the ${\cal X}(k_n)$ at $k_n = Q_F$ and when the peak diminishes ${\cal P}$ goes to zero. 2) The area (${\cal A}$) under the PMF curve (shown in \fig{fig:FFLO}) with respect to that of the $U=0$ case. It gives a measure of the pairing of particles with respect to the non-interacting case when there is no pairing. We show the variation of ${\cal P}$ in \fig{fig:FFLO_prop}(a) and of ${\cal A}$ in \fig{fig:FFLO_prop}(b) as a function of $U$. We note that both of them decreases with decreasing $U$ due to the suppression of the FFLO correlations noted in \fig{fig:FFLO_prop}(b). This suppression is stronger for smaller values of $\phi$, generating sharp decreases in ${\cal P}$ and ${\cal A}$ but for larger $\phi$, they change smoothly. Interestingly, we also note that the variation of ${\cal P}$ with $U$ for smaller values of $\phi$ is similar to that of an order parameter in standard phase transitions, i.~e. it is zero when this are no FFLO states while it becomes non-zero when FFLO states appear in the system. But, for larger values of flux, due to the presence of a peak even for the non-interacting case, as discussed earlier, there are smooth changes in both ${\cal P}$ and ${\cal A}$.

The suppression of the FFLO correlations is also related to the formation of two-body resonance like features in the scattering continuum as discussed in \sect{sec:two_body_results}. In the parameter regime, where these resonance like features appear, the state becomes a strongly interacting normal state. The FFLO correlations become short-ranged and are no longer dominant correlations of the system. It will be interesting to investigate different properties of this state and explore other quasi-long range orders in the system. Similar physics has also been pointed out in $3$D spin-orbit coupled Fermi gases with detuning and Zeeman field \cite{Shenoy2013}.

In \fig{fig:Corr_prop}, we show the behaviors of a local and a few non-local correlation functions of the $M=2$ SD system in real space. The local correlation function under consideration is the onsite average density of particle in the lowest leg $\langle n_{i,1} \rangle$. Its behavior is shown in \fig{fig:Corr_prop}(a) as a function of the site number $i$ for different values of $U$ and Friedel oscillations expected for a system with OBC \cite{Rizzi08} are seen. In its inset we show the difference in the onsite populations of the two legs $\Delta n_i = (\langle n_{i,1} \rangle - \langle n_{i,2} \rangle$) and see $\Delta n_i = 0$ for all values of $i$. From this figure, we stress the point that for the  parameter regime under consideration, there is no ``population imbalance'' in the system. Hence, these FFLO states are different from those predicted in the imbalanced $1$D Fermi gases \cite{Andreas2008,Feiguin2007,Rizzi08,Tezuka08,Feiguin2011} and are solely the effect of the synthetic gauge field present in the SD system (similar results of flow enhanced pairing are also seen in $3$D Fermi gases with SOC \cite{Shenoy2013}). Finally, in \fig{fig:Corr_prop}(b) we show the following non-local correlation functions with respect to the central site at $L/2$: a particular case of the PCF ${\cal X}_0 = {\cal X}_{i,L/2}$ (see \eqn{eqn:pcf}), single particle correlation function corresponding to the lowest ($\sigma = 1$) leg ${\cal C}_1 = \langle b^{\dagger}_{i,1} b_{L/2,1} \rangle$ and the highest ($\sigma = 2$) leg ${\cal C}_2 = \langle b^{\dagger}_{i,2} b_{L/2,2} \rangle$. We note that the single particle correlations are short-ranged but the PCF ${\cal X}_0$ shows algebraic decay with distance and is the slowest to decay. This signals the existence of a quasi-long-range order \cite{Feiguin2011} in the system with dominant FFLO correlations. 

\section{Summary and Outlook}
\mylabel{sec:summary}

In summary, we have investigated the interplay of the synthetic gauge field and an \SU{M} symmetric attractive interaction in the SD system. We showed that the synthetic gauge field changes the single particle spectrum of the SD system significantly and with a fixed density of particles this change leads to a Lifshitz transition of the Fermi surface from having $2$ Fermi points to $4$ Fermi points. We then focused on analyzing the novel effects brought solely by the synthetic gauge field and followed the didactic route of the BCS analysis by considering the two-body instabilities of the system first and then looking for their consequences in the many-body setting.

 Using the $T$-matrix formulation, we showed that the synthetic gauge field causes unusual effects on the two-body bound state spectrum of the system. It produces dimers having finite momentum which scales linearly with the magnetic flux. They can become two-body resonance like features in the scattering continuum with a large change in the phase shift with decreasing the strength of the interaction. As a result, even for this $1$D system a critical value of attraction strength is required to form bound states.

 Using DMRG, we then showed that these features give rise to exotic many body ground states such as the FFLO state. The FFLO states appear in the system even without any ``imbalance'' solely due to the synthetic gauge field present in the system in contrast to the usual $1$D Fermi gases with population imbalance. The FFLO momentum of the pairs formed in the system scales linearly with the magnetic flux. These states disappear gradually with continuous decrease in interaction strength and are present only above a critical value of interaction having similar behaviors as the two-body bound states. We have analyzed different properties of these states and showed that there are interesting measures to diagnose their presence in the system.

 On the other hand, we mentioned that a non-interacting fermionic SD system has already been experimentally realized in the ref. \cite{Mancini15} using $^{173}$Yb atoms. \SU{M} symmetric interaction can be produced by using orbital Feshbach resonances~\cite{Zhang2015,Hofer2015,Pagano2015} in this system. Also, there are other potential candidates for the experimental realizations of the \SU{M} symmetric fermionic SD systems such as using $^{6}$Li \cite{Bartenstein05} atoms. Finally, we would also like to point out that the SD system has the potential to realize a multi-flavor generalization of an interesting topological model known as the Creutz ladder \cite{Creutz94} model. This model has many interesting properties like the production of topological defects \cite{Bermudez09}, generation of persistent currents \cite{Sticlet13}, decay of edge states \cite{Viyuela12} etc. and show interesting behaviors in the presence of interactions \cite{Sticlet14}. Since the onsite spin-flip terms are already present in the SD system, the additional ingredient necessary for this realization is the generation of nearest neighbor spin-flip terms. These can be achieved by following the proposal of the ref.~\cite{Mazza2012} to induce controlled Raman transitions between nearest neighbor different flavor particles. We conclude by hoping that the interesting results presented in this article will be useful for further studies in this system. 

\paragraph*{Acknowledgements -} The authors acknowledge Vijay B. Shenoy for extensive discussions and comments on the manuscript. Adhip Agarwala is acknowledged for comments on the manuscript. The DMRG calculations have been performed using the DMRG code released within the “Powder with Power” Project (qti.sns.it).

\paragraph*{Author contributions -} UKY contributed to the formulation of the two-body physics only and SKG has done all the rest including the manuscript preparation.

\appendix

\section{$T$-matrix formulation}
\mylabel{sec:Tmatfor}
In this appendix, we give the details of the $T$-matrix formulation of the two-body problem of the SD system with \SU{M} symmetric attractive interaction discussed in the text. This general formulation accommodates both fermionic and bosonic particles and we use a parameter $\zeta$ which is $-1$ for fermions and $+1$ for bosons. To proceed, we first note that the pair creation operator (defined in \eqn{eqn:pair_create}) can be rewritten as
\beq\mylabel{eqn:pairpottemp}
P^{\dagger}_Q(\sigma,\sigma') = \halfsum{k} \left[ b^{\dagger}_{\frac{Q}{2}+k,\sigma} b^{\dagger}_{\frac{Q}{2}-k,\sigma'} + \zeta b^{\dagger}_{\frac{Q}{2}-k,\sigma'} b^{\dagger}_{\frac{Q}{2}+k,\sigma}\right].
\eeq
Now, we want to express this operator in terms of the two body state $\ket{K} = a^\dagger_{k_1,\alpha_1} a^\dagger_{k_2, \alpha_2} \ket{0}$ already defined in the text. To this end, we use the unitary matrices $R_{\sigma,\alpha_1}(k,k_1)$ and recast the above \eqn{eqn:pairpottemp} as
\bea
P^{\dagger}_Q(\sigma,\sigma') &=& \sum_{K} V^{Q}_{\sigma,\sigma'}(K)\ket{K} \,\,, \non\\
&=& \halfsum{K} W^{Q}_{\sigma,\sigma'}(K)\ket{K}\,\,.\mylabel{eqn:pairpot}
\eea
Here, we have defined 
\bea
V_{\sigma,\sigma'}^Q(K) &=& \halfsum{k} \bigg[R_{\sigma, \alpha_1}\bigg(\frac{Q}{2} + k, k_1\bigg) R_{\sigma', \alpha_2}\bigg(\frac{Q}{2} - k,k_2\bigg) \non\\
&+&  \zeta R_{\sigma', \alpha_1}\bigg(\frac{Q}{2} + k, k_1\bigg) R_{\sigma, \alpha_2}\left(\frac{Q}{2} - k,k_2\right)\bigg]\;, \non
\eea 

$\halfsum{k} \equiv \sum_{k \ge 0}$ and $W_{\sigma,\sigma'}^Q(K) = \left[V_{\sigma,\sigma'}^Q(K) + \zeta V_{\sigma,\sigma'}^Q(\tilde{K})\right]$. We note that $W_{\sigma,\sigma'}^Q(K)$ can be thought of as the potential felt by the two-body state $\ket{K}$ or the amplitude of the pair with COM $Q$ in the state $\ket{K}$. Denoting the scattering channels as $c \equiv (\sigma, \sigma')$, the interaction term ${\cal H}_I$ (\eqn{eqn:int_mom}) takes the form
\beq
{\cal H}_I = \sum_Q \halfsum{K,K'} \dW_Q(K,K') \ket{K} \bra{K'} \mylabel{eqn:interaction_eff}\,\,,
\eeq
where,
\beq\mylabel{eqn:eff_sca_pot}
\dW_Q(K,K') = -\frac{U}{2 L} \sum_c W_{c}^Q(K) W_{c}^{Q*}(K')\,\,,
\eeq
which can be thought of as the total effective scattering potential acting over all the $c$ scattering channels with fixed $Q$. As described in the text, there are $\kappa$ number of independent scattering channels in the system.

For a given $Q$, we now use the $T$-matrix formalism (closely following \cite{Taylor2006,Jayantha_twobody,Shenoy2013}) to write the $T$-matrix equation as (suppressing the $Q$ labels)
\beq\mylabel{eqn:tmat}
T_{K,K'}(\omega) = \dW(K,K') + \halfsum{K_1} \dW(K,K_1) G_0(K_1,\omega) T_{K_1,K'}(\omega),
\eeq
where, $G_0(K,\omega) = 1/(\omega^+ - E(K))$ is the two particle non-interacting Green's function and $\omega^+ \equiv (\omega + j 0)$. The above equation can be recast into the following form
\beq\mylabel{eqn:eff_tmat}
T_{K,K'}(\omega) = -\frac{U}{2L} \sum_c W_c(K) \left[W_c^{*}(K') + \Gamma_c(K', \omega)\right] \,\,,
\eeq
which reveals the fact that the $T$-matrix is separable in incoming and outgoing state contributions in each channel. Here, we have defined
\beq\mylabel{eqn:eff_Gamma}
\Gamma_c(K, \omega) = \halfsum{K_1} W_c(K_1) G_0(K_1,\omega) T_{K_1,K}(\omega) \;.
\eeq
Then using \eqn{eqn:eff_tmat} in the above \eqn{eqn:eff_Gamma}, we note that $\Gamma_c(K, \omega)$ satisfies the following equation,
\beq\mylabel{eqn:Gamma_eqn}
\mysum{c_1}\left[\delta_{c,c_1}+U\Lambda_{c,c_1}(\omega)\right]\Gamma_{c_1}(K,\omega) = U \mysum{c_1}\Lambda_{c,c_1}(\omega)W_{c_1}^{*}(K),
\eeq
with
\beq
\Lambda_{c,c_1}(\omega) = \frac{1}{2 L}\halfsum{K} W_c^{*}(K) G_0(K,\omega) W_{c_1}(K)\,\,.
\eeq
We now define two column vectors $\utilde{\Gamma}(K,\omega)$ and $\utilde{W}(K)$ whose $c$-th elements are $\Gamma_{c}(K,\omega)$ and $W^*_c(K)$ respectively. And, we also define two matrices $\utilde{\Lambda}(\omega)$ and the all important secular matrix $\utilde{\dL}(\omega)$ whose $(c,c_1)$-th elements are $\Lambda_{c,c_1}(\omega)$ and $L_{c,c_1}(\omega) = \left[ \delta_{c,c_1} + U \Lambda_{c,c_1}(\omega) \right]$ respectively. Then, we can solve the 
\eqn{eqn:Gamma_eqn} formally to obtain
\beq
\utilde{\Gamma}(K,\omega) = U \utilde{\dL}^{-1}(\omega).\utilde{\Lambda}(\omega).\utilde{W}(K) \,\,.
\eeq
Plugging this equation in \eqn{eqn:eff_tmat}, we note that the poles of the $T$-matrix, which give the energies of the $\kappa$ number of bound states, are determined by the solutions of the equation
\beq\mylabel{eqn:inveqn}
{\cal D}et\left[\utilde{\dL}(\omega)\right] = 0 \,\,.
\eeq
For a general $\phi$, therefore, we solve this equation numerically and obtain different bound state properties of the system.

\bibliography{refFFLO}

\end{document}